\documentclass[english,draftcls,onecolumn,11pt]{IEEEtran}
\usepackage[T1]{fontenc}
\usepackage[latin9]{inputenc}
\usepackage{color}
\usepackage{array}
\usepackage{multirow}
\usepackage{amsmath}
\usepackage{amsthm}
\usepackage{amssymb}
\usepackage{setspace}

\makeatletter

\providecommand{\tabularnewline}{\\}

  \theoremstyle{definition}
  \newtheorem{defn}{\protect\definitionname}
\theoremstyle{plain}
\newtheorem{thm}{\protect\theoremname}
  \theoremstyle{plain}
  \newtheorem{cor}{\protect\corollaryname}
 \theoremstyle{definition}
  \newtheorem{example}{\protect\examplename}

\usepackage{amsfonts}
\usepackage{algorithmic}
\usepackage{algorithm}
\usepackage{multirow}
\usepackage{cite}
\usepackage{xcolor}

\newtheorem{ctexample}{Counter-Example}

\makeatother

\usepackage{babel}
  \providecommand{\definitionname}{Definition}
  \providecommand{\examplename}{Example}
\providecommand{\corollaryname}{Corollary}
\providecommand{\theoremname}{Theorem}

\begin{document}

\title{ExpSOS: Secure and Verifiable Outsourcing of Exponentiation Operations
for Mobile Cloud Computing}

\author{Kai Zhou, M. H. Afifi and Jian Ren\thanks{The authors are with the Department of Electrical and Computer Engineering,
Michigan State University, East Lansing, MI 48824-1226, Email: \{zhoukai,
afifi, renjian\}@msu.edu}}
\maketitle
\begin{abstract}
Discrete exponential operation, such as modular exponentiation and
scalar multiplication on elliptic curves, is a basic operation of
many public-key cryptosystems. However, the exponential operations
are considered prohibitively expensive for resource-constrained mobile
devices. In this paper, we address the problem of secure outsourcing
of exponentiation operations to one single untrusted server. Our proposed
scheme (ExpSOS) only requires very limited number of modular multiplications
at local mobile environment thus it can achieve impressive computational
gain. ExpSOS also provides a secure verification scheme with probability
approximately $1$ to ensure that the mobile end-users can always
receive valid results. The comprehensive analysis as well as the simulation
results in real mobile device demonstrates that our proposed ExpSOS
can significantly improve the existing schemes in efficiency, security
and result verifiability. We apply ExpSOS to securely outsource several
cryptographic protocols to show that ExpSOS is widely applicable to
many cryptographic computations.\end{abstract}


\section{Introduction}

Cloud computing provides end-users the capability to securely access
the shared pool of resources such as computational power and storage.
It enables end-users to utilize those resources in a pay-per-use manner.
Among all types of computations, exponential operation in a finite
group is almost ubiquitous in public-key cryptosystems. However, due
to large integers involved, exponentiation is considered prohibitively
expensive for resource-constrained devices such as mobile phones.
Thus, outsourcing exponentiation operation to the cloud servers becomes
an appealing choice. 

However, when sensitive data is outsourced to the untrusted cloud,
security of the data as well as the result is at risk. Moreover, many
cryptographic applications, such as digital signature, require to
verify the validity of the results of modular exponentiation. Thus
result verification is also a crucial issue. In contrast, the cloud
cannot be fully trusted for at least three reasons. First, the cloud
could be curious. That is, it may try to ``mine'' as much information
as possible from the outsourced data. Second, the computational resource
is commodity. The cloud has the motivation to cheat in the computation
process in order to save computational resources. Third, the cloud
is a shared environment. It is hard to secure individual data using
just regular processor. Thus, security and verifiability are two major
concerns for computation outsourcing.

To address these two issues, various computation outsourcing mechanisms
have been proposed, including outsourcing of modular exponentiation
operations \cite{atallah2002,atallah2010,atallah2005secure,wang2011infocom,chen2014efficient,blanton2012secure,blanton2010secure,hohenberger2005securely,chen2012new,wang2014securely}.
In \cite{hohenberger2005securely}, the authors considered outsourcing
modular exponentiation to two servers assuming that they would not
collude. The basic idea of the proposed scheme in \cite{hohenberger2005securely}
is to split the base and exponent of modular exponentiation into random
looking pieces that are separately outsourced to two servers. Then
the end-user can combine the results returned by the servers to recover
the desired result. Under this scheme, the end-user can check the
validity of the returned results with probability $\frac{1}{2}$.
Following \cite{hohenberger2005securely}, the authors in \cite{chen2012new}
proposed a similar scheme and improved the performance by reducing
one query to the servers and increasing the verifiability to $\frac{2}{3}$.
In order to eliminate the assumption that the two servers would not
collude, the authors in \cite{wang2014securely} proposed a scheme
to outsource modular exponentiation to one single server. However,
at local side, the end-user still needs to carry out some exponentiation
operations. As a result, the computational gain is limited for the
end-user. Moreover, all these three schemes rely on pre-computation
of modular exponentiation of some random integers. This will cause
extra overhead to end-user's limited computational power or storage
space depending on the method by which pre-computation is implemented.

From the above analysis of several previous schemes, we can summarize
some basic requirements of secure outsourcing of modular exponentiation.
First, for the system model, it is much more desirable to outsource
exponentiation operations to one single server instead of two servers
with security based on the assumption that two servers would not collude.
Second, the secure outsourcing scheme should not impose expensive
computational overhead at local side. Otherwise, the performance gain
from outsourcing would diminish. Third, the scheme should provide
a high verifiability. Ideally, the end-user should be able to verify
the validity of the returned result with probability $1$.

In this paper, we extend the notion of exponentiation from modular
exponentiation to general exponential operations in a finite group,
including scalar multiplication on elliptic curves. In general, each
exponential operation consists of a series of basic group operations.
The number of such operations varies with the exponent. In this sense,
modular exponentiation and scalar multiplication can both be regarded
as exponentiation operations. Thus, we propose a Secure Outsourcing
Scheme for general Exponential (ExpSOS) operations. The proposed ExpSOS
is based on ring homomorphism. Specifically, we map the integers in
the ring $\mathbb{R}_{N}$ to the ring $\mathbb{R}_{L}$ so that the
computation in $\mathbb{R}_{L}$ is homomorphic to that in $\mathbb{R}_{N}$.
We let the cloud carry out the computation in $\mathbb{R}_{L}$ and
from the result returned by the cloud, the end-user is able to recover
the result back to $\mathbb{R}_{N}$ efficiently. The ring homomorphism
has two features: i) the mapping between $\mathbb{R}_{N}$ and $\mathbb{R}_{L}$
is computationally efficient, and ii) without possessing the secret
key, it is computationally infeasible to derive any key information
of the result in $\mathbb{R}_{N}$ from that in $\mathbb{R}_{L}$.
The main contributions of this paper can be summarized as follows:
\begin{itemize}
\item We formally define a secure outsourcing scheme and four outsourcing
models. The proposed ExpSOS is shown to be effective under all four
different models.
\item We develop schemes to securely outsource exponentiation operations
in a general finite group, including modular exponentiation and scalar
multiplication on elliptic curves.
\item We outsource exponential operation to one single untrusted server
eliminating the non-collusion assumption between multiple servers.
\item Our proposed ExpSOS is efficient in that it requires only a small
number of modular multiplications at local side. 
\item We propose a verification scheme such that the end-user can verify
the validity of the result with probability approximately $1$.
\end{itemize}
The rest of this paper is organized as follows. In Section \ref{sec:Secure-Outsourcing-Model},
we introduce four secure outsourcing models and formally define a
secure outsourcing scheme. In Section \ref{sec:Proposed-Scheme},
we present the design of ExpSOS for both modular exponentiation and
scalar multiplication based on ring homomorphism. We propose the verification
scheme in Section \ref{sec:Result-Verification}. The complexity and
security analysis of ExpSOS are given in Section \ref{sec:Complexity-and-Security}.
Then we apply ExpSOS to outsource several cryptographic protocols
in Section \ref{sec:Application}. In Section \ref{sec:Performance-Comparison},
we compare the performance of ExpSOS with several existing works and
give some numeric results. We conclude in Section \ref{sec:Conclusion}.

\section{Secure Computation Outsourcing Model\label{sec:Secure-Outsourcing-Model}}

\subsection{System Model and Threat Model}

\paragraph{System Model}

In the general settings of computation outsourcing, the system consists
of two entities: an end-user $E$ and the cloud $S$. The end-user
$E$ is resource-constrained. It has limited computational power and
storage space. The cloud $S$ is regarded as possessing abundant resources
and is able to carry out expensive computations. The cloud can be
further modeled as the \textit{single-server} model and the \textit{multiple-servers}
model. In the single-server model, the cloud is viewed as one unit.
In contrast, in the multiple-servers model, the cloud is divided into
two or more individual units. Each unit carries out the computational
tasks independently. While communication between different units is
allowed, key information is only limited to individual unit since
otherwise security of the whole system maybe in jeopardy. 

Suppose the end-user $E$ wishes to accomplish a computationally expensive
task $F(\mathbf{x})\rightarrow\omega$, where $\mathbf{x}$ is the
input and $\omega$ is the output of the task. However, due to the
limited resources, $E$ may not be able to finish the task using the
locally available resources. The computational task $F$ could be
outsourced to $S$. Unfortunately, the cloud is only a shared server
and cannot be fully trusted. Therefore, we have to make sure that
it is infeasible for $S$ to derive any key information about both
$\mathbf{x}$ and $\omega$ from the outsourced task.

\paragraph{Threat Model}

We propose two threat models for the cloud. First, the cloud $S$
is \textit{honest but curious}. That is, the cloud will honestly fulfill
its advertised functionality. However, $S$ could be curious. It may
try to exploit any key information from the outsourced task, which
may include the input, the output as well as the intermediate computational
results. When the outsourced data is sensitive, this could cause severe
security and privacy issues. Second, the cloud $S$ is \textit{malicious},
meaning that the cloud $S$ may not carry out the desired computation
truthfully. This can happen for various reasons. A simple scenario
could be that the cloud simply returns some trivial results since
the computational resource is a commodity for the cloud server. As
a consequence, the end-user $E$ is unable to receive a valid result
from the cloud server $S$. 

Based on the above system model and threat model, we can divide the
computation outsourcing scenarios into four types in a hierarchical
manner: 
\begin{itemize}
\item \textbf{MS}: Malicious cloud under Single-server model. 
\item \textbf{HCS}: Honest but Curious cloud under Single-server model.
\item \textbf{MM}: Malicious cloud under Multiple-servers model.
\item \textbf{HCM}: Honest but Curious cloud under Multiple-servers model. 
\end{itemize}
It is hierarchical in the sense that a secure outsourcing scheme designed
for single-server model can be extended to multiple-servers model
and a scheme for malicious cloud can be extended to honest but curious
cloud. Specifically, these four models can be organized into three
layers: at the bottom layer is the HCM model, in the middle are the
MM and HCS and on the top is MS. A secure outsourcing scheme designed
for a model in an upper layer is also suitable for that in a lower
layer . Thus, a secure outsourcing scheme for MS is most widely applicable
and achieves the highest security standard. In this paper, we first
propose a secure outsourcing scheme for the HCS model. Then a verification
scheme is proposed for MS model.

\subsection{Definition of Secure Outsourcing Scheme}

A secure computation outsourcing scheme mainly addresses two issues:
the security of the outsourced computational problem and the validity
of the returned results. We formally define a Secure Outsourcing Scheme
(SOS) as a 4-tuple $(\mathcal{T},\mathcal{C},\mathcal{R},\mathcal{V})$
consisting of four different functions:
\begin{enumerate}
\item \textbf{Problem Transformation} $\mathcal{T}:F(\mathbf{x})\to G(\mathbf{y})$.
The end-user $E$ locally transforms the problem $F(\mathbf{x})$
to a new form $G(\mathbf{y})$, where $\mathbf{y}$ is the new input
and $G$ is the new problem description. $E$ then outsources $G(\mathbf{y})$
to the cloud server $S$.
\item \textbf{Cloud Computation} $\mathcal{C}:G(\mathbf{y})\rightarrow(\Omega,\Gamma)$.
The cloud $S$ solves the transformed problem $G(\mathbf{y})$ to
obtain the corresponding result $\Omega$. At the same time, $S$
returns $\Gamma$ that is a proof of the validity of the result.
\item \textbf{Result Recovery} $\mathcal{R}:\Omega\rightarrow\omega$. Based
on the returned result $\Omega$, the end-user $E$ recovers the result
$\omega$ of the original problem $F(\mathbf{x})$.
\item \textbf{Result Verification} $\mathcal{V}:(\Omega,\Gamma,\omega)\rightarrow\top=\{\mathsf{True},\mathsf{False}\}$.
Based on $\omega,\Omega$ and the proof $\Gamma$, the end-user $E$
verifies the validity of the result.
\end{enumerate}
An SOS should satisfy the following two requirements:
\begin{enumerate}
\item \textbf{Soundness}: given that the cloud is honest but curious, $E$
can successfully recover the correct result $\omega$ from the returned
result $\Omega$. That is $\mathcal{R}(\Omega)=\omega$.
\item \textbf{Security}: the cloud is unable to derive any key information
about the original input $\mathbf{x}$ and output $\omega$ from the
transformed problem $G$, the new input $\mathbf{y}$ and the new
output $\Omega$. 
\end{enumerate}
To measure the performance of an SOS, we adopt a similar definition
of efficiency and verifiability as proposed in \cite{hohenberger2005securely}.
We introduce the following two definitions:
\begin{defn}[$\alpha$-efficient]
\label{def:efficient} Suppose the running time of a task $F$ for
$E$ is $t_{0}$. Under an SOS, the running time of local processing
for $E$ is $t_{p}$. Then the SOS is $\alpha$-efficient if $\frac{t_{0}}{t_{p}}\geq\alpha$.
\end{defn}

\begin{defn}[$\beta$-verifiable]
\label{def:checkable} Given the returned output $\Omega$ and the
proof $\Gamma$, denote the probability that $E$ is able to verify
the validity of the result $\omega$ as $\rho$. Then an SOS is $\beta$-verifiable
if $\rho\geq\beta$.
\end{defn}
From the definition above, we can see that a larger $\alpha$ indicates
a better performance of a secure outsourcing scheme, while a larger
$\beta$ means a better verifiability.

\section{Secure Outsourcing of Exponentiation Operations \label{sec:Proposed-Scheme}}

In this section, we first define a ring homomorphism $f:\:\mathbb{R}_{1}\rightarrow\mathbb{R}_{2}$.
Based on this ring homomorphism, we propose a secure outsourcing scheme
for exponentiation operations. In this section, the threat model is
assumed to be HCS.\textcolor{red}{{} }However, our proposed verification
scheme ensures that ExpSOS is secure under the MS model.

\subsection{Ring Homomorphism}

Consider two rings and their corresponding operations $(\mathbb{R}_{1},+,\cdot)$
and $(\mathbb{R}_{2},\circ,\star)$ and a mapping function $f:\;\mathbb{R}_{1}\rightarrow\mathbb{R}_{2}$.
We define ring homomorphism as follows:
\begin{defn}[\textbf{Ring Homomorphism}]
 Given $(\mathbb{R}_{1},+,\cdot)$ and $(\mathbb{R}_{2},\circ,\star)$,
a mapping function $f:\;\mathbb{R}_{1}\rightarrow\mathbb{R}_{2}$
is a ring homomorphism if there exists an inverse mapping function
$g:\;\mathbb{R}_{2}\rightarrow\mathbb{R}_{1}$ and the pair $(f,g)$
possesses the following two properties:\end{defn}
\begin{itemize}
\item \textbf{Additive Homomorphism}: $\forall x_{1},x_{2}\in\mathbb{R}_{1}$,
$x_{1}+x_{2}=g(f(x_{1})\circ f(x_{2}))$;
\item \textbf{Multiplicative Homomorphism}: $\forall x_{1},x_{2}\in\mathbb{R}_{1}$,
$x_{1}\cdot x_{2}=g(f(x_{1})\star f(x_{2}))$.
\end{itemize}
In this paper, we assume that exponentiation operations are operated
in the ring $\mathbb{R}_{N}$. We note that $N$ is not necessarily
a prime. It can also be product of large primes. Then, our primitive
goal is to construct a proper ring homomorphism $f:\mathbb{R}_{N}\to\mathbb{R}_{L}$
that maps elements in $\mathbb{R}_{N}$ to elements in another ring
denoted as $\mathbb{R}_{L}$. In this way, the computations in $\mathbb{R}_{N}$
can be concealed when transformed to the corresponding computations
in $\mathbb{R}_{L}$ so that the computations in $\mathbb{R}_{N}$
can be concealed. 

Define $f:\mathbb{R}_{N}\to\mathbb{R}_{L}$ as follows:

\begin{equation}
f(x)=\begin{array}{c}
(x+kN)\bmod L,\end{array}\label{eq:ring-homomorphism}
\end{equation}
where $k$ is a random integer in $\mathbb{R}_{N}$, $L=pN$ and $p$
is a large prime. The following theorem states that the proposed $f$
achieves ring homomorphism.
\begin{thm}
\label{thm:ring-homo}$\forall x\in\mathbb{R}_{N}$, the mapping $f$
defined in equation (\ref{eq:ring-homomorphism}) is a ring homomorphism.\end{thm}
\begin{IEEEproof}
We show that there exists an inverse mapping function $g:\mathbb{R}_{L}\to\mathbb{R}_{N}$
and the pair $(f,g)$ possesses both the additive and the multiplicative
homomorphic properties. Define the inverse mapping function $g$ as
\[
\begin{array}{c}
g(y)=y\bmod N.\end{array}
\]

Suppose $x_{1},x_{2}\in\mathbb{R}_{N}$, $f(x_{1})=(x_{1}+k_{1}N)\bmod L$
and $f(x_{2})=(x_{2}+k_{2}N)\bmod L$, where $k_{1},k_{2}\in\mathbb{R}_{N}$
are randomly selected integers. We can verify that 

\[
\begin{array}{cl}
 & g(f(x_{1})+f(x_{2}))\\
= & ((x_{1}+k_{1}N)\bmod L+(x_{2}+k_{2}N)\bmod L)\bmod N\\
= & (x_{1}+k_{1}N+x_{2}+k_{2}N)\bmod L\bmod N\\
= & (x_{1}+k_{1}N+x_{2}+k_{2}N)\bmod N\\
= & (x_{1}+x_{2})\bmod N.
\end{array}.
\]

Thus, we have proved that $(f,g)$ has additive homomorphic property.
Similarly, we can verify that $(f,g)$ is also multiplicative homomorphic
as follows:

\[
\begin{array}{cl}
 & g(f(x_{1})\cdot f(x_{2}))\\
= & ((x_{1}+k_{1}N)\bmod L\cdot(x_{2}+k_{2}N)\bmod L)\bmod N\\
= & ((x_{1}+k_{1}N)\cdot(x_{2}+k_{2}N))\bmod L\bmod N\\
= & ((x_{1}+k_{1}N)\cdot(x_{2}+k_{2}N))\bmod N\\
= & x_{1}\cdot x_{2}\bmod N.
\end{array}.
\]

Hence, the proposed mapping function $f(x)=(x+kN)\bmod L$ is a ring
homomorphism.
\end{IEEEproof}
The above proposed ring homomorphism enables us to transform the addition
and multiplication in a ring into the corresponding operations in
another large ring. We further explore the polynomial homomorphic
property of the ring homomorphism that is defined as follows.
\begin{defn}[\textbf{Polynomial Homomorphism}]
 Suppose $\mathbf{x}=(x_{1},x_{2},\cdots,x_{n})\in\mathbb{R}_{N}^{n}$
and ${\rm poly}(\mathbf{x})$ is a polynomial function defined on
$\mathbf{x}$. A mapping function $f:\;\mathbb{R}_{N}\longrightarrow\mathbb{R}_{L}$
is polynomial homomorphic if there exists an inverse mapping function
$g:\;\mathbb{R}_{L}\longrightarrow\mathbb{R}_{N}$ such that 
\[
g({\rm poly}(f(\mathbf{x})))={\rm poly}(\mathbf{x}),
\]
where $f$ is applied on $\mathbf{x}$ opponent-wise. \end{defn}
\begin{thm}
\label{thm:The-proposed-ring}The proposed ring homomorphism $f(x)=(x+kN)\bmod L$
is polynomial-homomorphic.
\end{thm}
The proof of the above theorem is straightforward given the additive
and multiplicative homomorphic properties of the ring homomorphism.

\subsection{ExpSOS under HCS Model}

In this section, we will consider two kinds of exponentiation operations,
that are modular exponentiation and scalar multiplication on elliptic
curves.

\subsubsection{Secure Outsourcing of Modular Exponentiation}

Consider modular exponentiation $R=u^{a}\bmod N$. We assume that
$N$ is either a large prime or a product of large prime numbers,
which is the typical situation in cryptosystems. Theorem \ref{thm:ring-homo}
states that the result of multiplication in the ring $\mathbb{R}_{N}$
can be obtained from the multiplication in $\mathbb{R}_{L}$ through
the transformation function and the inverse function. If we take $x_{1}=x_{2}=u$,
we can get
\[
((u+rN)\bmod L)^{2}\bmod N=u^{2}\bmod N.
\]
If we repeat the multiplication in $\mathbb{R}_{N}$ for $a$ times,
we have the following corollary.
\begin{cor}
\label{cor:exponentiation} For $u,a,r\in\mathbb{R}_{N}$, we have
\[
((u+rN)\bmod L)^{a}\bmod N=u^{a}\bmod N.
\]

\end{cor}
Corollary \ref{cor:exponentiation} gives us a way to conceal the
base when outsourcing modular exponentiation. That is, we can first
transform the original base $u$ to $U=(u+rN)\bmod L$, where $r\in\mathbb{R}_{N}$
is a random integer. Then the cloud can compute $U^{a}\bmod L$ based
on which the result can be recovered by computing $(U^{a}\bmod L)\bmod N=u^{a}\bmod N$.
As long as $N$ is kept secret, the cloud cannot learn the value of
$u$ due to the randomness of $r$. 

The remaining task is to conceal the exponent $a$. We have the following
theorem.
\begin{thm}
\label{thm:euler} For $N=p_{1}p_{2}\cdots p_{m}$, where $p_{1},p_{2},\cdots,p_{m}$
are distinct prime numbers, we have
\[
u^{a+k\phi(N)}\bmod N=u^{a}\bmod N,
\]
where $k$ is a random integer and $\phi(\cdot)$ is the Euler's totient
function.\end{thm}
\begin{IEEEproof}
We first prove $u^{1+k\phi(N)}\bmod N=u\bmod N$. Consider a prime
factor $p_{i}$ of $N$, $i=1,2,\cdots,m$. There are two cases:
\begin{itemize}
\item Case 1: $\mbox{\ensuremath{\gcd}}(u,p_{i})\neq1$, that is $u$ and
$p_{i}$ are not relatively prime. In this case, we have $p_{i}\mid u$.
Thus
\[
(u^{1+k\phi(N)}-u)\bmod p_{i}=0,
\]
which means that $p_{i}\mid(u^{1+k\phi(N)}-u)$.
\item Case 2: $\gcd(u,p_{i})=1$, that is $u$ and $p_{i}$ are relatively
prime. Then, by the Euler's Theorem, we have $u^{\phi(p_{i})}\bmod p_{i}=1$.
From the multiplicative property of the Euler's totient function,
we have $\phi(N)=\phi(p_{1})\phi(p_{2})\cdots\phi(p_{m})$. Let $\theta(p_{i})=\phi(N)/\phi(p_{i})$.
Then,
\[
\begin{array}{ll}
 & u^{1+k\phi(N)}\bmod p_{i}\\
= & u\cdot u^{k\phi(p_{1})\phi(p_{2})\cdots\phi(p_{m})}\bmod p_{i}\\
= & u\bmod p_{i}\cdot(u^{\phi(p_{i})}\bmod p_{i})^{k\theta(p_{i})}\bmod p_{i}\\
= & u\bmod p_{i}\cdot(1)^{k\theta(p_{i})}\bmod p_{i}\\
= & u\bmod p_{i}.
\end{array}
\]

That is $(u^{1+k\phi(N)}-u)\bmod p_{i}=0$.

Thus, in both cases, we have proved that $p_{i}\mid(u^{1+k\phi(N)}-u).$
Since $p_{i}$ is arbitrarily selected and $p_{1},p_{2},\cdots,p_{m}$
are distinct primes, we have
\[
N\mid(u^{1+k\phi(N)}-u).
\]

Hence, $u^{1+k\phi(N)}\bmod N=u\bmod N$. Multiplying both sides of
the equation by $u^{a-1}$, we can obtain
\[
u^{a+k\phi(N)}\bmod N=u^{a}\bmod N.
\]

\end{itemize}
\end{IEEEproof}
In Theorem \ref{thm:euler}, we do not require that $u$ and $N$
to be co-prime as required in the Euler's theorem. Instead, we assume
that $N$ is the product of distinct primes that is typical in cryptosystems.
For instance, in RSA, the modulus $N=pq$ is the product of two distinct
prime numbers.

Theorem \ref{thm:euler} introduces a way to conceal the exponent
$a$. That is, by transforming the original exponent $a$ to $A=a+k\phi(N)$,
where $k$ is a random integer, we can conceal $a$ due to the randomness
of $k$. Now, based on Theorem \ref{thm:ring-homo} and Theorem \ref{thm:euler},
we can construct our secure outsourcing scheme for modular exponentiation.
In the secure outsourcing scheme, the function $\mathcal{C}(U,A,L)$
outsourced to the could can be expressed as a modular exponentiation
$\mathcal{C}(U,A,L)=U^{A}\bmod L$. The result recovery function is
$\mathcal{R}(R,N)=R\bmod N$. The secure outsourcing scheme for modular
exponentiation under HCS model is given in Algorithm \ref{alg:ExpSOS-under-HCS}. 

\begin{algorithm}[tbh] 
\caption{Secure Outsourcing of Modular Exponentiation Under HCS Model\label{alg:ExpSOS-under-HCS}}

\smallskip 
\textbf{Input:} $N,u,a\in\mathbb{R}_N$.\\
\textbf{Output:} $R=u^a\bmod N$.

\smallskip 
$\mathsf{Key\ Generation}$:
\begin{algorithmic}[1] 
\STATE $E$ generates a large prime $p$ and calculate $L\leftarrow pN$. 
\STATE The public key is $K_p=\{L\}$, and the private key is $K_s=\{p,N\}$.  
\end{algorithmic} 

\smallskip 
$\mathsf{Problem\ Transformation}\ \mathcal{T}$:  
\begin{algorithmic}[1] 
\STATE $E$ selects random integers $r,k\in\mathbb{R}_N$ as the temporary key.
\STATE $E$ calculates $A\leftarrow a+k\phi(N)$,
$U\leftarrow (u+rN)\bmod L$.
\STATE $E$ outsources $\mathcal{C}(U,A,L)$ to the cloud.
\end{algorithmic} 

\smallskip
$\mathsf{Cloud\ Computation}\ \mathcal{C}$:
\begin{algorithmic}[1] 
\STATE $S$ computes $R_1\leftarrow\mathcal{C}(U,A,L)=U^A\bmod L$. 
\STATE $S$ returns $R_1$ to $E$.
\end{algorithmic} 

\smallskip
$\mathsf{Result\ Recovery}\ \mathcal{R}$:
\begin{algorithmic}[1] 
\STATE $E$ recovers the result as $R\leftarrow \mathcal{R}(R_1)=R_1\bmod N$.
\end{algorithmic} 
\end{algorithm}

The soundness of the outsourcing scheme is guaranteed by the following
theorem:
\begin{thm}
\label{thm:correctness} The secure outsourcing scheme for modular
exponentiation is sound. That is $R=R_{1}\bmod N=u^{a}\bmod N$.
\end{thm}
The proof of Theorem \ref{thm:correctness} is straightforward based
on Theorem \ref{thm:ring-homo} and Theorem \ref{thm:euler}. Specifically,
by transforming the original problem of modular exponentiation to
a disguised form, our proposed ExpSOS under HCS model is sound.

\subsubsection{Secure Outsourcing of Scalar Multiplication}

In this section, we consider secure outsourcing of scalar multiplication
$sP$ on an elliptic curve $E(\mathbb{F}_{p})$ described by the following
short Weierstrass equation: 

\begin{equation}
E:\;y^{2}=x^{3}+bx+c,\label{eq:Short}
\end{equation}
where the coefficients $b,c$ and the coordinates of the points are
all in a finite field $\mathbb{F}_{p}$. Furthermore, for cryptographic
applications, we usually work with points in a set of $m$-torsion
points $E(F_{p})[m]$ defined as $E(F_{p})[m]=\{P\in E(F_{p}):[m]P=\mathcal{O}\}$,
where $\mathcal{O}$ is the point at infinity. Thus, we assume $P\in E(F_{p})[m]$
and $s\in\mathbb{Z}_{m}$.

The secure outsourcing of scalar multiplication relies on two basic
operations that are point addition and point doubling. They play a
similar role as modular multiplication in the outsourcing of modular
exponentiation. Specifically, the ``double-and-add'' algorithm to
calculate scalar multiplication on elliptic curves consists of a series
of point addition and point doubling. Thus intuitively, we can regard
secure outsourcing of point addition and point doubling as two building
blocks to implement scalar multiplication.

We utilize projective coordinate to represent a point $P=(x,y,z)$
corresponding to the point $Q=\left(\frac{x}{z},\frac{y}{z}\right)$
in the affine coordinates. As a result, the computation of point addition
and point doubling consists of only modular addition and multiplication.
Specifically, given two points $P=(x_{1},y_{1},z_{1})$ and $Q=(x_{2,}y_{2},z_{2})$
such that $P\neq\pm Q$, the point addition $P+Q=(x_{3},y_{3},z_{3})$
can be calculated as follows:
\[
x_{3}=BC,y_{3}=A(B^{2}x_{1}z_{2}-C)-B^{3}y_{1}z_{2},z_{3}=B^{3}z_{1}z_{2},
\]
where 
\[
\begin{array}{l}
A=y_{2}z_{1}-y_{1}z_{2},B=x_{2}z_{1}-x_{1}z_{2},\\
C=A^{2}z_{1}z_{2}-B^{3}-2B^{2}x_{1}z_{2}.
\end{array}
\]
The point doubling $2P=(x_{4},y_{4},z_{4})$ can be calculated as
follows:
\[
x_{4}=2BD,y_{4}=A(4C-D)-8y_{1}^{2}B^{2},z_{4}=8B^{3},
\]
where 
\[
A=bz_{1}^{2}+3x_{1}^{2},B=y_{1}z_{1},C=x_{1}y_{1}B,D=A^{2}-8C.
\]
In projective coordinates, one point addition and doubling take $14$
multiplications and $12$ multiplications, respectively. 

Theorem \ref{thm:The-proposed-ring} states that by mapping the variables
of a polynomial from a finite field to variables in a ring, we can
evaluate the polynomial in the ring and recover the result in the
finite field. This gives us the insight of our proposed scheme since
essentially, point addition and point doubling are both the process
of evaluating polynomials on the coordinates of the points. Thus,
we can construct the secure computation scheme for point addition
and point doubling as in Algorithm \ref{alg:Secure-Point-Addition}. 

\begin{algorithm}[tbh] 
\caption{Secure Point Addition and Point Doubling\label{alg:Secure-Point-Addition}}

\textbf{Input:} $P=(x_1,y_1,z_1)$, $Q=(x_2,y_2,z_2)$ and $E=\{b,c,p \}$.\\
\textbf{Output:} point $R=P+Q=(x_3,y_3,z_3)$.

\begin{algorithmic}[1] 
\STATE Select a large prime $p$ and compute $N=pq$. 
\STATE For a coordinate $x_i$, select a random integer $k_i$ and compute $x_i'=(x_i+k_ip)\bmod N$.
\STATE Transform the points $P,Q$ and the elliptic curve $E$ to $P'=(x_1',y_1',z_1')$, $Q'=(x_2',y_2',z_2')$ and $E'=\{b',c',N\}$ respectively as described in Step 2. 
\STATE Outsource $P', Q'$ and $E'$ to the cloud.
\STATE Cloud computes $R'=P'+Q'$ following the point doubling or point addition prodecure.
\STATE On receiving $R'=(x_3',y_3',z_3')$, recover $R$ as $R=(x_3',y_3',z_3')\bmod p=(x_3,y_3,z_3)$.
\end{algorithmic} 
\end{algorithm}
\begin{thm}
\label{thm:The-proposed-secure}The proposed secure point addition
and point doubling algorithm is sound.
\end{thm}
The proof of Theorem \ref{thm:The-proposed-secure} is straightforward
from the polynomial-homomorphic property of the ring homomorphism.

The above theorem enables us to conceal the points as well as the
parameters of the elliptic curve from the cloud. To outsource scalar
multiplication $sP$, the remaining part is to conceal the multiplier
$s$. We utilize the property of the order $m$ of the torsion group
that is $rmP=\mathcal{O}$, for an arbitrary point $P\in E[m](\mathbb{F}_{p})$
and any integer $r$. As a result, we can conceal $s$ by adding it
to a multiple of $m$ as $s^{\prime}=s+rm$, where $r$ is a random
integer. Now, we can summarize the secure outsourcing scheme of scalar
multiplication as in Algorithm \ref{alg:Secure-Outsourcing}.

\begin{algorithm}[tbh]
\caption{Secure Outsourcing of Scalar Multiplication Under HCS Model\label{alg:Secure-Outsourcing}}

\textbf{Input:} $P=(x_1,y_1,z_1)$, $s$, $E=\{b,c,p\}$ and $m$.\\
\textbf{Output:} point $R=sP$.

$\mathsf{Key~Generation}$:
\begin{algorithmic}[1] 
\STATE End-user selects a large prime $q$ and compute $N\leftarrow pq$.
\end{algorithmic} 

$\mathsf{Problem~Transformation}$:
\begin{algorithmic}[1] 
\STATE End-user generates random integers $k_1,k_2,k_3,k_4,k_6,r$.
\STATE Computes $x_1'\leftarrow (x_1+k_1p)\bmod N$, $y_1'\leftarrow (y_1+k_2p)\bmod N$, $z_1'\leftarrow (z_1+k_3p)\bmod N$, $b'\leftarrow (b+k_4p)\bmod N$, $c'\leftarrow (c+k_6p)\bmod N$, $s'\leftarrow s+rm$. 
\STATE End-user outsources $P'=(x_1',y_1',z_1')$, $E'=\{ b',c',N\} $ and $s'$.
\end{algorithmic} 

$\mathsf{Cloud~Computation}$:
\begin{algorithmic}[1] 
\STATE The cloud computes $R'\leftarrow s'P'$ utilizing the double-and-add algorithm.
\end{algorithmic}

$\mathsf{Result~Recovery}$:
\begin{algorithmic}[1] 
\STATE The end-user recovers the result $R$ as $R\leftarrow (x_3', y_3',z_3' )\bmod p$.
\end{algorithmic} 
\end{algorithm}
\begin{thm}
The secure outsourcing scheme for scalar multiplication is sound.
That is $R=sP$.\end{thm}
\begin{IEEEproof}
From Theorem \ref{thm:The-proposed-secure}, we know that the secure
computation scheme for point addition and point doubling is sound.
Since the double-and-add algorithm to compute scalar multiplication
consists of a series of point addition and point doubling, we have
$R=s^{\prime}P=(s+rm)P=sP+rmP=sP+\mathcal{O}=sP$.
\end{IEEEproof}
In the next section, we propose a verification scheme to ensure that
ExpSOS is secure under the MS model.

\section{Result Verification \label{sec:Result-Verification}}

In this section, we first analyze the necessary properties of a result
verification scheme through some counter examples. We then propose
a result verification scheme for the outsourcing of modular exponentiation
under MS model. We show that the verification scheme can also be applied
to the outsourcing of scalar multiplication.

In the HCS model discussed in the previous section, we assume that
the cloud will honestly conduct its advertised functionality. That
is, to compute the function $\mathcal{C}(U,A,L)$ and return the correct
result $U^{A}\bmod L$. However, in the MS model, the cloud may manipulate
the result in order to save computational resources. Thus, to verify
the soundness of the result returned by the cloud is a critical issue.

A natural way to verify the result, as utilized in many previous works
\cite{chen2012new,chen2014efficient,hohenberger2005securely}, is
to outsource the problem multiple times and verify whether the returned
results satisfy certain criteria. However, this methodology may cause
potential security problems if it is not carefully designed. This
is because outsourcing multiple times essentially gives more information
about the original problem to the cloud, which may increase the probability
for the cloud to recover the original problem. Moreover, the cloud
may manipulate the results in order to satisfy the criteria, thus
passing the verification. Therefore, we believe that an effective
verification scheme should at least have the following two properties: 
\begin{itemize}
\item \textbf{Security}: The verification process should not reveal any
key information about the original problem to the cloud. 
\item \textbf{Anti-manipulation}: It is infeasible for the cloud to manipulate
the result and pass the verification process. 
\end{itemize}
We utilize two counter-examples in verifying modular exponentiation
to illustrate the significance of the above properties and emphasize
the key issues in designing a verification scheme. 

\begin{ctexample}
Transform the exponent $a$  to $A_1=a+k_{1}\phi(N)$ and $A_2=a+k_{2}\phi(N)$. The cloud returns results $R_1=U^{A_{1}}\bmod L$ and $R_2=U^{A_2}\bmod L$. The end-user checks whether the condition  $R_1\bmod N=R_2\bmod N$ holds.  
\end{ctexample}

Unfortunately, the above example violates the security property. When
the cloud possesses $A_{1}$ and $A_{2}$, it can calculate $A_{1}-A_{2}=(k_{1}-k_{2})\phi(N)$,
which is a multiple of the Euler's totient function $\phi(N)$. In
this case, the cloud can factorize $(k_{1}-k_{2})\phi(N)$ based on
which, the cloud may be able to check the primality of $N$. Since
$N$ is a product of large primes, the consequence is that the cloud
can limit the valid value of $N$ to a short list. That is the cloud
have a good chance to guess the value of $N$. This means that the
cloud can derive some key information from the outsourced problem
thus making outsourcing insecure. Similarly, some variances of this
type of method (e.g., $A_{1}=a+k_{1}\phi(N)$ and $A_{2}=ca+k_{2}\phi(N)$,
where $c$ is a known constant) may also have security problems.

\begin{ctexample}
Transform the exponent $a$ to $A_1=a+k_1\phi(N)$ and $A_2=a+t+k_2\phi(N)$, where $t$ is a relatively small integer and calculating $u^{t}\bmod N$ is within the end-user's computational ability. The cloud returns results $R_1=U^{A_1}\bmod L$ and $R_2=U^{A_2}\bmod L$. The end-user checks whether the condition $(R_1\cdot u^t)\bmod N=R_2\bmod N$ holds.
\end{ctexample}

Due to the randomness of $t$, the cloud is not able to obtain a multiple
of $\phi(N)$. However, from the equality condition $(R_{1}\cdot u^{t})\bmod N=R_{2}\bmod N$,
we have $U^{A_{1}}\cdot u^{t}\bmod N=U^{A_{2}}\bmod N$, which is
equivalent to 

\[
u^{t}\bmod N=U^{A_{2}-A_{1}}\bmod N.
\]

In this case, the cloud can manipulate two arbitrary integers $A_{1}^{\prime}$
and $A_{2}^{\prime}$ as long as $A_{2}^{\prime}-A_{1}^{\prime}=A_{2}-A_{1}$.
The results will pass the verification but the recovered result $R=U^{A_{1}^{\prime}}\bmod N$
is incorrect. This means that the cloud can manipulate a false result
while passing the verification process.

From the above two counter examples, we can see that security and
anti-manipulation are two critical issues in result verification schemes.
In the following Algorithm \ref{alg:Result-Verification-Scheme},
we propose a verification scheme for modular exponentiation.

\begin{algorithm}[tbh] 
\caption{ExpSOS under MS Model\label{alg:Result-Verification-Scheme}}

\smallskip 
\textbf{Input:} $N,u,a\in\mathbb{R}_N$.\\
\textbf{Output:} $R_0=u^a\bmod N$, $\Lambda=\{\mathsf{True},\mathsf{False}\}.$

\smallskip 
$\mathsf{Key~Generation}$:
\begin{algorithmic}[1] 
\STATE $E$ generates a large prime $p$ and calculate $L\leftarrow pN$. 
\STATE The public key is $K_p=\{L\}$, and the private key is $K_s=\{p,N\}$.  
\end{algorithmic} 

\smallskip 
$\mathsf{Problem~Transformation}~\mathcal{T}$:  
\begin{algorithmic}[1] 
\STATE $E$ selects random integers $r,k_1,k_2,t_1,t_2$  as the ephemeral key with the constraint that $t_1,t_2\leq b$.
\STATE $E$ calculates $A_1\leftarrow a+k_1\phi(N)$, $A_2 \leftarrow t_1 a+t_2+k_2 \phi(N)$ and $U\leftarrow (u+rN)\bmod L$.
\STATE $E$ outsources $\mathcal{C}(U,A_1,L)$ and $\mathcal{C}(U,A_2,L)$ to the cloud.
\end{algorithmic} 

\smallskip
$\mathsf{Cloud~Computation}~\mathcal{C}$:
\begin{algorithmic}[1] 
\STATE $S$ computes $R_1\leftarrow\mathcal{C}(U,A_1,L)\leftarrow U^{A_1}\bmod L$ and $R_2\leftarrow\mathcal{C}(U,A_2,L)\leftarrow U^{A_2}\bmod L$.
\STATE $S$ returns $R_1$ and $R_2$ to $E$.
\end{algorithmic} 

\smallskip
$\mathsf{Result~Verification}~\mathcal{V}$:
\begin{algorithmic}[1] 
\STATE $E$ checks $(R_1 \bmod N)^{t_1} \cdot u^{t_2} \bmod N = R_2 \bmod N$.
\STATE If the equality holds, set $\Lambda \leftarrow \mathsf{True}$. Otherwise, set $\Lambda \leftarrow \mathsf{False}$.
\end{algorithmic} 

\smallskip
$\mathsf{Result~Recovery}~\mathcal{R}$:
\begin{algorithmic}[1] 
\STATE $E$ recovers the result as $R_0\leftarrow \mathcal{R}(R_1)=R_1\bmod N$.
\end{algorithmic} 
\end{algorithm}

Now, we utilize an example to illustrate our proposed ExpSOS under
MS model.
\begin{example}
Suppose the end-user $E$ wants to calculate $u^{a}\bmod N$, where
$N=431$ is a prime, $u=189$ and $a=346$. $E$ can outsource $u^{a}\bmod N$
as follow:\end{example}
\begin{enumerate}
\item $\mathsf{Key\ Generation}$: $E$ select a prime number $p=397$ and
calculate $L=pN=171107$. Then $E$ selects random integers $r=146,k_{1}=332,k_{2}=68$
and $t_{1}=4,t_{2}=12$ with $t_{1},t_{2}<b=16$.
\item $\mathsf{Problem\ Transformation}$: $E$ calculates $A_{1}=a+k_{1}\phi(N)=143106$,
$A_{2}=t_{1}a+t_{2}+k_{2}\phi(N)=30636$ and $U=(u+rN)\bmod L=63115$.
$E$ then queries $\mathcal{C}(U,A_{1},L)$ and $\mathcal{C}(U,A_{2},L)$
to the cloud $S$.
\item $\mathsf{Cloud\ computation}$: $S$ computes $R_{1}=U^{A_{1}}\bmod L=63115^{143106}\bmod171107=81281$,
$R_{2}=U^{A_{2}}\bmod L=63115^{30636}\bmod171107=55473$ and returns
$R_{1}$ and $R_{2}$ to $E$.
\item $\mathsf{Result\ Verification}$: $E$ calculates $(R_{1}\bmod N)^{t_{1}}\cdot u^{t_{2}}\bmod N=(190^{4}\cdot189^{12})\bmod431=305$
and $R_{2}\bmod N=55473\bmod431=305$ that satisfy $(R_{1}\bmod N)^{t_{1}}\cdot u^{t_{2}}\bmod N=R_{2}\bmod N$.
Thus the returned results are correct.
\item Result Recovery: $E$ recovers the result as $R=R_{1}\bmod N=81281\bmod431=190$
that is equal to $u^{a}\bmod N=190$.
\end{enumerate}
In Algorithm \ref{alg:Result-Verification-Scheme}, the two outsourced
exponential operations are related through an affine function. As
a consequence, the cloud is unable to derive a multiple of $\phi(N)$
only based on $A_{1}$ and $A_{2}$. Moreover, the cloud cannot manipulate
the results to create a verifiable equality. 

This verification scheme can also be applied to the outsourcing of
scalar multiplications. The base point $P$ can be transformed to
$P^{\prime}$ as described in Algorithm \ref{alg:Secure-Outsourcing}.
The exponent $s$ can be transformed to $s_{1}=s+r_{1}m$ and $s_{2}=t_{1}s+t_{2}+r_{2}m$,
where $r_{1},r_{2},t_{1},t_{2}$ are random integers and $t_{1},t_{2}\leq b$.
Then the end-user can check the condition $Q_{2}=t_{1}Q_{1}+t_{2}P$,
where $Q_{1}=s_{1}P^{\prime}$ and $Q_{2}=s_{2}P^{\prime}$.

\section{Complexity and Security Analysis \label{sec:Complexity-and-Security}}

In this section, we analyze the security and the computational complexity
of ExpSOS. We utilize the secure outsourcing of modular exponentiation
as a representative to perform the analysis. The analysis of outsourcing
scalar multiplication can be conducted in a similar way. We show that
ExpSOS is secure under both HCS and MS model. Specifically, under
the HCS model, the ExpSOS is $\frac{1}{2}\log_{2}a$-efficient. Under
the MS model, the ExpSOS is $\frac{1}{2}\log_{b}a$-efficient and
$(1-\frac{1}{2b^{2}})$-verifiable, where $a$ is the exponent and
$b$ is the security parameter.

\subsection{Security Analysis}

In ExpSOS, we conceal the base $u$ through a ring homomorphism $(u+rN)\bmod L$
and the exponent $a$ is mapped to $a+k\phi(N)$. In our analysis,
we show that given the public information $\{L,U,A_{1},A_{2}\}$,
the cloud cannot derive any key information about the input $\{u,a,p\}$
and the output $R=u^{a}\bmod p$.

First, the following theorem shows that the ring homomorphism is secure.
\begin{thm}
\label{thm:ring-security} When the integers $N$ and $p$ are sufficiently
large, it is computationally infeasible to recover $u$ from the ring
homomorphism $f:\:u\mapsto U=(u+rN)\bmod L$.\end{thm}
\begin{IEEEproof}
The security is based on the hardness of integer factorization. That
is, given $L=pN$, where $p$ and $N$ are large prime numbers, it
is computationally infeasible to factorize $L$ to get $p$ and $N$.
In our case, we consider the module $N$ as a large prime number or
a product of large prime numbers, which is typical in cryptosystems.
Thus, given $L$, the cloud is unable to recover $N$. Furthermore,
as $r$ is a random integer, given $U=(u+rN)\bmod L$, the cloud is
also unable to recover $u$. \end{IEEEproof}
\begin{thm}
In the ExpSOS scheme, it is computationally infeasible to recover
the exponent $a$ under both HCS and MS model.\end{thm}
\begin{IEEEproof}
The proof is straightforward since under the HCS model, the cloud
obtains $A=a+k\phi(N)$, while under the MS model, the cloud obtains
$A_{1}=a+k_{1}\phi(N)$ and $A_{2}=t_{1}a+t_{2}+k_{2}\phi(N)$. In
both cases, the randomness of $k,k_{1},k_{2},t_{1},t_{2}$ and security
of the totient function $\phi(N)$ make it infeasible for the cloud
server to derive the exponent $a.$
\end{IEEEproof}
We show that the proposed verification scheme has the security and
effectiveness properties as described previously. First, the security
is based on the likelihood of finding two integers $R_{1}$ and $R_{2}$
so that $(R_{1}\bmod N)^{t_{1}}\cdot u^{t_{2}}\bmod N=R_{2}\bmod N$
holds true, and deriving a multiple of $\phi(N)$ from $A_{1}=a+k_{1}\phi(N)$,
and $A_{2}=t_{1}a+t_{2}+k_{2}\phi(N)$. The former would enable the
cloud server to cheat the end-user without conducting the actual computation
and the latter could make it possible for the cloud server to recover
$\phi(N)$ and then perform collision attacks. 
\begin{thm}
For any two randomly selected integers $R_{1}$ and $R_{2},$ the
probability that $(R_{1}\bmod N)^{t_{1}}\cdot u^{t_{2}}\bmod N=R_{2}\bmod N$
is $1/b^{2}.$ \end{thm}
\begin{IEEEproof}
The proof of this theorem is straightforward since only one pair of
$(t_{1},t_{2})$ will make the equality holds true, while the total
number of possible combinations for the $(t_{1},t_{2})$ pair is $b^{2}.$ 
\end{IEEEproof}
This theorem indicates that if the cloud wants to manipulate the result,
it has to guess the random integers, the probability to succeed is
only $1/b^{2}$. In fact, if we outsource $\mathcal{C}(U,A_{1},L)$
and $\mathcal{C}(U,A_{2},L)$ in a random order, we can further reduce
the probability for the cloud to guess the correct randoms to $1/(2b^{2}).$
According to Definition \ref{def:checkable}, ExpSOS is at least $(1-1/(2b^{2}))$-verifiable. 
\begin{thm}
For any two randomly selected integer $t_{1}$ and $t_{2},$ the probability
to derive a multiple of $\phi(N)$ is at most $1/b^{2}.$ \end{thm}
\begin{IEEEproof}
Since $A_{1}=a+k_{1}\phi(N)$ and $A_{2}=t_{1}a+t_{2}+k_{2}\phi(N)$,
and $t_{1}$ is a randomly chosen integer from $(0,b],$ the cloud
server has probability $1/b$ to get the right $t_{1}$ and derive
the following equation
\begin{equation}
(t_{1}k_{1}-k_{2})\phi(N)=(t_{1}A_{1}-A_{2})+t_{2},\label{eq:Verification-Collision}
\end{equation}
where $A_{1}$ and $A_{2}$ are known and $t_{1},t_{2}$ are secretly
selected. For the right-hand side of this equation, if further $t_{2}$
is known, then its integer factorization could potentially reveal
the factors of $\phi(N).$ However, since $t_{2}$ is randomly chosen
in the range $(0,b]$, the likelihood to get a proper $t_{2}$ is
$1/b.$ Therefore, the overall probability to obtain equation (\ref{eq:Verification-Collision})
is $1/b^{2}.$
\end{IEEEproof}
The upper bound $b$ is a security parameter that measures the confidence
of the end-user about the returned result. In practical computation
outsourcing systems, the cloud would be severely punished if cloud
manipulation is detected. Therefore, the benefit for the cloud to
cheat would be hardly justifiable in this setting.

\subsection{Complexity Analysis}

We utilize outsourcing of modular exponentiation as a representative
to analysis complexity. The analysis can be applied to scalar multiplication
similarly. The essence of ExpSOS is to limit the number of modular
multiplications for the end-user to compute modular exponentiation
with the aid of the cloud. In our analysis, we utilize the number
of modular multiplications, denoted as $\pi$, as a measurement. To
calculate $u^{a}\bmod N$, the number of multiplications is $\pi=\frac{3}{2}l_{a}$,
where $l_{a}$ is the bit length of $a$\cite{zhong2000modular}.
Therefore, in calculating the modular exponentiation $u^{a}\bmod N$,
$l_{a}\approx\log_{2}a$ and $\pi\approx\frac{3}{2}\log_{2}a$.

In ExpSOS, under the HCS model, to calculate $U,A$ and $L$, the
end-user needs $3$ multiplications. We notice that when the end-user
knows the factors of $N$, it is computationally easy to calculate
$\phi(N)$. For example, when $N$ is a prime, $\phi(N)=N-1$. Moreover,
the calculation of $\phi(N)$ is a one-time process. The computational
overhead for calculating $\phi(N)$ is negligible especially when
the end-user outsources modular exponentiation multiple times. Thus,
under HCS model, we have $\pi_{HCS}=3$. Hence, the computational
gain from outsourcing is $\alpha_{HCS}=\pi/\pi_{HCS}=\frac{1}{2}\log_{2}a$.
From Definition \ref{def:efficient}, ExpSOS is $\frac{1}{2}\log_{2}a$-efficient
under the HCS model. 

Under the MS model, the calculation of $L,U,A_{1},A_{2}$ will take
$4$ multiplications. In the verification scheme, the end-user has
to calculate $(R_{1}\bmod N)^{t_{1}}\bmod N$ and $u^{t_{2}}\bmod N$.
Thus, $\pi_{MS}=4+\frac{3}{2}\log_{2}t_{1}+\frac{3}{2}\log_{2}t_{2}+1$.
Since $t_{1}$ and $t_{2}$ are upper-bounded by $b$, we have $\log_{2}t_{1}+\log_{2}t_{2}\leq2\log_{2}b$.
Hence the computational gain from outsourcing is 
\[
{\displaystyle \begin{array}{rcl}
\alpha & = & {\displaystyle \frac{\pi}{\pi_{MS}}}\\
 & = & {\displaystyle \frac{\frac{3}{2}\log_{2}a}{5+\frac{3}{2}\log_{2}t_{1}+\frac{3}{2}\log_{2}t_{2}}}\\
 & \geq & {\displaystyle \frac{\frac{3}{2}\log_{2}a}{5+3\log_{2}b}}\\
 & \approx & {\displaystyle \frac{1}{2}\log_{b}a}.
\end{array}}
\]
 Thus under the MS model, ExpSOS is at least $\frac{1}{2}\log_{b}a$-efficient.

\subsection{Trade-Off between Computation and Security}

The above security and complexity analysis reveal the trade-off between
computational overhead and security. In the MS model, ExpSOS is at
least $\frac{1}{2}\log_{b}a$-efficient and $(1-1/(2b^{2}))$-verifiable.
Both measurements relate to the same parameter $b$. On one hand,
$b$ is the upper bound of the computational overhead that the end-user
can tolerate. On the other hand, $b$ reveals the confidence of the
end-user about the returned result which is also regarded as the security
level of the result. When $b$ increases, the end-user has to carry
out more computation. However, the probability that the end-user can
verify the validity of the result also increases. 

Thus, the proposed ExpSOS is cost-aware in the sense that it enables
the end-user to have the flexibility to choose the most suitable outsourcing
scheme according to its computational constraint and security demand.
This is important especially when the end-users vary in computational
power and security demands. It also makes ExpSOS widely applicable.

\section{Applications\label{sec:Application}}

The proposed ExpSOS is able to conceal the base, the exponent and
the module of the modular exponentiation $u^{a}\bmod N$. It can also
be used to conceal the base point $P$ and multiplier $s$ of the
scalar multiplication $sP$. With this feature, the parameters (private
or public) within the cryptosystem are totally concealed from the
outside especially the cloud. Thus, the cryptosystem is isolated from
the outsourced system. In this sense, ExpSOS can be regarded as a
black box that takes as input $\{u,a,N,b\}$ and creates the output
$u^{a}\bmod N$ as $\mathsf{ExpSOS}(u,a,N,b)\rightarrow u^{a}\bmod N$,
where $b$ is security parameter selected by the end-user. The end-user
will have a performance gain of $\frac{1}{2}\log_{b}a$ and can verify
the validity of the result with probability $1-\frac{1}{2b^{2}}$. 

In this section, we will explore efficient outsourcing of exponential
operations in some typical cryptographic protocols to the cloud. We
will first introduce the outsourcing of Digital Signature Algorithm
(DSA) that involves only modular exponentiation. Then, we illustrate
how to outsource the encryption part of Identity Based Encryption
(IBE) system involving both modular exponentiation and scalar multiplication.

\subsection{Outsourcing DSA Operations}

We utilize DSA \cite{william2003} as an example of digital signature
schemes. In DSA, the global public key component $\{p,q,g\}$ is shared
by a group of users. Here, $p,q$ are prime numbers and $q$ is a
divisor of $p-1$. $g=h^{(p-1)/q}\bmod p$ with $1<h<(p-1)$ such
that $h^{(p-1)/q}\bmod p>1$. The algorithm can be divided into the
following three phases:
\begin{enumerate}
\item \textbf{Key Generation}: The signer $E$ generates a private key $x$
with $0<x<q$ and calculates the public key as $y=g^{x}\bmod p$.
\item \textbf{Signing}: $E$ selects a private key $k$ with $0<k<q$ and
calculates $r=(g^{k}\bmod p)\bmod q$, $s=(k^{-1}(h(M)+xr))\bmod q$,
where $M$ is the message and $h(M)$ is the hash value of $M$ using
SHA-1. The signature of $M$ is $\{r,s\}$.
\item \textbf{Verifying}: A verifier $V$ calculates $\omega=s^{-1}\bmod q$,
$u_{1}=(h(M)\omega)\bmod q$, $u_{2}=rw\bmod q$ and $v=(g^{u_{1}}y^{u_{2}})\bmod p\bmod q$.
Then the verifier checks whether $v=r$ is true.
\end{enumerate}

\begin{algorithm}[h] 
\caption{Secure Outsourcing of DSA message signing\label{alg:SecureDSA-Sign}}

\smallskip 
$\mathsf{Key~Generation}$:  
\begin{algorithmic}[1] 
\STATE $E$ selects a large prime number $Q$ and calculate $L\leftarrow Qp$. 
\end{algorithmic} 

\smallskip 
$\mathsf{Problem~Transformation}~\mathcal{T}$:  
\begin{algorithmic}[1] 
\STATE $E$ selects temporary key $r_1,k_1,k_2,k_3,t_1,t_2,t_3$ with $t_1,t_2,t_3<b$. 
\STATE $E$ calculates $X\leftarrow x+k_1\phi(p)$, $K\leftarrow k+k_2\phi(p)$, $X_K\leftarrow t_1x+t_2k+t_3+k_3\phi(p)$ and $G\leftarrow (g+r_1p)\bmod L$.
\STATE $E$ outsources $\mathcal{C}(X,G,L)$, $\mathcal{C}(K,G,L)$ and $\mathcal{C}(X_K,G,L)$ in random order to the cloud $S$.
\end{algorithmic} 

\smallskip
$\mathsf{Cloud~Computation}~\mathcal{C}$:
\begin{algorithmic}[1] 
\STATE $S$ computes $R_1\leftarrow G^X\bmod L$, $R_2\leftarrow G^K\bmod L$ and $R_3\leftarrow G^{X_K}\bmod L$.
\STATE $S$ returns the results $R_1$, $R_2$ and $R_3$ to $E$.
\end{algorithmic} 

\smallskip
$\mathsf{Result~Verification}~\mathcal{V}$:
\begin{algorithmic}[1] 
\STATE $E$ verifies the results by checking $((R_1\bmod p)^{t_1}\cdot(R_2\bmod p)^{t_2}\cdot g^{t_3}\bmod p)\bmod p=R_3\bmod p$.
\end{algorithmic} 

\smallskip
$\mathsf{Result~Recovery}~\mathcal{R}$:
\begin{algorithmic}[1] 
\STATE $E$ recovers the results $y\leftarrow R_1\bmod p$ and $r\leftarrow (R_2\bmod p)\bmod q$.
\end{algorithmic} 

\smallskip
$\mathsf{Signature~Generation}$: 
\begin{algorithmic}[1] 
\STATE $E$ generates the signature $\{r,s\}$ by calculating $s\leftarrow (k^{-1}(h(M)+xr))\bmod q$. 
\STATE $E$ shares the public information $\{G,R_1,L\}$ within the group of users.
\end{algorithmic}
\end{algorithm}

\begin{algorithm}[h] 
\caption{Secure Outsourcing of DSA sigature verification\label{alg:SecureDSA-Verification}}

\smallskip 
$\mathsf{Problem~Transformation}~\mathcal{T}$:  
\begin{algorithmic}[1] 
\STATE The verifier $V$ generates temporary key $k_4,k_5,k_6,k_7,t_4,t_5,t_6,t_7$ with $t_4,t_5,t_6,t_7<b$.
\STATE $V$ calculates $U_1\leftarrow u_1+k_4\phi(p)$, $U_2\leftarrow u_2+k_5\phi(p)$, $U_3\leftarrow t_4u_1+t_5+k_6\phi(p)$ and $U_4\leftarrow t_6u_2+t_7+k_7\phi(p)$. 
\STATE $V$ outsources $\mathcal{C}(G,U_1,L)$, $\mathcal{C}(G,U_2,L)$, $\mathcal{C}(R_1,U_3,L)$ and $\mathcal{C}(R_1,U_4,L)$ to the
cloud.
\end{algorithmic} 

\smallskip
$\mathsf{Cloud~Computation}~\mathcal{C}$:
\begin{algorithmic}[1] 
\STATE $S$ calculates $R_4\leftarrow G^{U_1}\bmod L$, $R_5\leftarrow G^{U_2}\bmod L$, $R_6\leftarrow R_1^{U_3}\bmod L$, $R_7\leftarrow R_1^{U_4}\bmod L$ 
\STATE $S$ returns the results $\{R_4,R_5,R_6,R_7\}$ to $V$.
\end{algorithmic} 

\smallskip
$\mathsf{Result~Verification}~\mathcal{V}$:
\begin{algorithmic}[1] 
\STATE $V$ verifies the results by checking $((R_4\bmod p)^{t_4}\cdot g^{t_5}\bmod p)\bmod p=R_6\bmod p$ and $((R_5\bmod p)^{t_6}\cdot(R_1\bmod p)^{t_7})\bmod p=R_7\bmod p$.
\end{algorithmic} 

\smallskip
$\mathsf{Result~Recovery}~\mathcal{R}$:
\begin{algorithmic}[1] 
\STATE $V$ recovers the results $g^{u_1}\bmod p\leftarrow R_4\bmod p$ and $y^{u_2}\bmod p\leftarrow R_6\bmod p$.
\end{algorithmic} 

\smallskip
$\mathsf{Signature~Verification}$: 
\begin{algorithmic}[1] 
\STATE $V$ calculates $v\leftarrow (g^{u_1}y^{u_2})\bmod p$ and check $v=r$.
\end{algorithmic}
\end{algorithm}

We can see that the computational bottleneck of DSA is the calculation
of $g^{x}\bmod p$, $g^{k}\bmod p$ for the signer and $(g^{u_{1}}y^{u_{2}})\bmod p$
for the verifier. We formulate the outsourcing of DSA in Algorithms
\ref{alg:SecureDSA-Sign} and Algorithm \ref{alg:SecureDSA-Verification}.
To outsource the two exponentiation operations $g^{x}\bmod p$, $g^{k}\bmod p$,
the signer $S$ makes $3$ queries to the cloud and carries out $\pi_{E}=(8+\frac{9}{2}\log b)$
modular multiplications. In comparison, the original computational
burden is $\pi_{0}=\frac{3}{2}(\log x+\log k)$. For the verifier
$V$, the computational overhead becomes $\pi_{V}=(6+6\log b)$ in
comparison with the original $\pi_{0}=\frac{3}{2}(\log u_{1}+\log u_{2})$.

\subsection{Outsourcing Identity Based Encryption}

Identity Based Encryption (IBE) system is proposed to alleviate the
process of public key certification in traditional public key cryptosystems.
In IBE system, a user can utilize his identity such as his email address\textit{
}as the public key. Then a trusted authority will generate and distribute
private key to the message receiver. The idea of IBE was initialized
by Shamir in \cite{shamir1985identity}. A practical IBE system was
proposed in \cite{boneh2001identity} based on bilinear pairing on
elliptic curves.

In an implementation of IBE system \cite[Chapter 5]{hoffstein2008introduction},
the public parameters are an elliptic curve $E(\mathbb{F}_{p})[m]$
and a base point $P\in E(\mathbb{F}_{p})[m]$. Also, the trusted authority
will publish his own public key $P_{T}\in E(\mathbb{F}_{p})[m]$.
The parameters are known to the authenticated users in the system.
We assume that a user Alice uses the hash of her own identity to generate
the public key which is a point on the elliptic curve, that is $P_{A}\in E(\mathbb{F}_{p})[m]$.
For any other user Bob who desires to send a message $M$ to Alice,
he will conduct the following encryption process:
\begin{enumerate}
\item Bob selects a random integer $r\in Z_{m}$;
\item Bob computes $C_{1}=rP$;
\item Bob computes $C_{2}=M\textsf{ \ensuremath{\oplus}\ }H(e(P_{A},P_{T}))^{r}$;
\item Bob sets the cipher text as $C=(C_{1},C_{2})$.
\end{enumerate}
In the above encryption algorithm, $e(P_{A},P_{T})$ denotes the pairing
between public points $P_{A}$ and $P_{T}$ and $H(\cdot)$ is a hash
. We note that both the input and output of the pairing $e(P_{A},P_{T})$
are public. Thus, the end-user Bob can obtain the pairing result denoted
as $g=e(P_{A},P_{T})$. To this end, we can see that the computational
burden for Bob lies in the scalar multiplication $rP$ and the modular
exponentiation $g^{r}\bmod p$. We summarize the outsourcing of IBE
as in Algorithm \ref{alg:SecureIBE}.

\begin{algorithm}[tbh] 
\caption{Secure Outsourcing of Identity Based Encryption\label{alg:SecureIBE}}

$\mathbf{Input}$: $P=(x,y,z)$, $r$, $g=e(P_A,P_T)$\\
$\mathbf{Output}$: $C_1=rP$, $C_2 = H(g)^r$

$\mathsf{Key~Generation}$:  
\begin{algorithmic}[1] 
\STATE Bob selects a large prime $q$ and calculates $L\leftarrow pq$. 
\end{algorithmic} 
\smallskip 
$\mathsf{Problem~Transformation}~\mathcal{T}$:  
\begin{algorithmic}[1] 
\STATE Bob generates temporary key $k_1, k_2, k_3, k_4, k_5, t_1,t_2$ with $t_1,t_2<b$.
\STATE Bob calculates $r_1\leftarrow (r+k_1 p)\bmod L$, $r_2\leftarrow (t_1r+t_2+k_2 p)\bmod L$, $x'\leftarrow (x+k_3 p)\bmod L$, $y'\leftarrow (y+k_4 p)\bmod L$, $z'\leftarrow (z+k_5 p)\bmod L$. Bob sets $P' \leftarrow (x',y',z')$.
\STATE Bob outsources $\mathcal{C}(r_1,P',E')$, $\mathcal{C}(r_2,P',E')$, $\mathcal{C}(r_1,H(g),L)$ and $\mathcal{C}(r_2,H(g),L)$ to the
cloud, where $E'$ is the transformed elliptic curve.
\end{algorithmic} 

\smallskip
$\mathsf{Cloud~Computation}~\mathcal{C}$:
\begin{algorithmic}[1] 
\STATE $S$ calculates $Q_1 \leftarrow r_1 P'$, $Q_2 \leftarrow r_2 P'$, $R_1 \leftarrow H(g)^{r_1}$ and $R_2 \leftarrow H(g)^{r_2}$.  
\STATE $S$ returns the results $\{Q_1, Q_2, R_1, R_2\}$ to Bob.
\end{algorithmic} 

\smallskip
$\mathsf{Result~Verification}~\mathcal{V}$:
\begin{algorithmic}[1] 
\STATE Bob verifies the results by checking $((R_1\bmod p)^{t_1}\cdot H(g)^{t_2}\bmod p)\bmod p=R_2\bmod p$ and $ (t_1 Q_1 + t_2 P)\bmod p =Q_2 \bmod p$, where the modular is applied coordinate-wise.
\end{algorithmic} 

\smallskip
$\mathsf{Result~Recovery}~\mathcal{R}$:
\begin{algorithmic}[1] 
\STATE Bob  recovers the results $C_1 \leftarrow Q_2\bmod p$ and $C_2 \leftarrow M  \oplus R_2\bmod p$.
\end{algorithmic} 
\end{algorithm}

From the above two applications, we can summarize some techniques
in designing secure outsourcing scheme utilizing the outsourcing of
exponential operation as a building block.
\begin{itemize}
\item It is more efficient and secure to share some common parameters in
different subroutines of the outsourcing process. For example, in
outsourcing of DSA, the signer and verifier share the same disguised
base $G$ and $R_{1}$. The benefits are that on one hand, the computational
overhead is reduced; on the other hand, less information is exposed
to the cloud. 
\item When outsourcing modular exponentiation with the same base, the computational
overhead can be reduced by jointly verifying the result. For example,
in outsourcing of of the DSA, the results of $g^{x}\bmod p$ and $g^{k}\bmod p$
can be jointly verified by constructing a common exponent $X_{K}=t_{1}x+t_{2}k+t_{3}+k_{3}\phi(p)$
that is a linear combination of the two disguised exponents $X$ and
$K$. Therefore, the signer does not have to carry out the extra exponentiation.
\item When making multiple queries to the cloud, the end-user can randomize
the order of queries to increase verifiability. For example, in outsourcing
of DSA, the signer and the verifier need to make $3$ and $4$ queries
to the cloud, respectively. If the order of queries are randomized,
the cloud has to guess the correct orders before guessing the correct
parameters. As a result, the verifiability for the signing process
increases to $1-\frac{1}{6b^{3}}$ and that of the verifying process
increases to $1-\frac{1}{24b^{4}}$. 
\end{itemize}

\section{Performance Evaluation\label{sec:Performance-Comparison}}

To the best of our knowledge, previous research on secure outsourcing
of cryptographic computations mainly focuses on modular exponentiation.
In this section, we first compare ExpSOS with three existing works
on secure outsourcing of modular exponentiation. Then we give some
numeric results to show the efficiency of ExpSOS.

\subsection{Performance Comparison}

\begin{table*}
\caption{Performance Comparison\label{tab:Performance-Comparison-1}}

\begin{spacing}{1.1}
\centering{}%
\begin{tabular}{|c|c|c|c|c|c|c|}
\hline 
\textbf{Scheme} & \textbf{Model} & \textbf{Pre-Processing} & \textbf{Multiplication} & \textbf{Inversion} & \textbf{Queries to Server} & \textbf{verifiability}\tabularnewline
\hline 
\hline 
\cite{hohenberger2005securely} & MM & $6$ $\mathsf{Rand}$ & $6$ $\mathcal{O}(\mathsf{Rand})+9$  & $5$ & $8$ & $1/2$\tabularnewline
\hline 
\cite{chen2012new} & MM & $5$ $\mathsf{Rand}$  & $5$ $\mathcal{O}(\mathsf{Rand})+7$ & $3$ & $6$ & $2/3$\tabularnewline
\hline 
\cite{wang2014securely} & MS & $7$ $\mathsf{Rand}$  & $7$ $\mathcal{O}(\mathsf{Rand})+\frac{3}{2}\log\chi+12$ & $4$ & $4$ & $1/2$\tabularnewline
\hline 
\multirow{3}{*}{ExpSOS} & HCS & Not Required & $3$ & $0$ & $1$ & Not Applicable\tabularnewline
\cline{2-7} 
 & MM & Not Required & $3$ & $0$ & $2$ & $1$\tabularnewline
\cline{2-7} 
 & MS & Not Required & $5+3\log b$ & $0$ & $2$ & $1-1/2b^{2}\approx1$\tabularnewline
\hline 
\end{tabular}\end{spacing}
\end{table*}

Secure outsourcing of cryptographic computations, especially modular
exponentiation, has been a popular research topic \cite{hohenberger2005securely,chen2012new,wang2014securely,matsumoto1990speeding,de1997schnorr,boyko1998speeding,nguyen2001distribution,van2006speeding}.
For instance, the authors in \cite{van2006speeding} proposed a secure
outsourcing scheme for modular exponentiation with variable-exponent
fixed base and fixed-exponent variable-base under single untrusted
server model. However, the base is known to the server. In \cite{hohenberger2005securely},
the authors considered outsourcing variable-base variable-exponent
modular exponentiation to two untrusted servers. Following this work,
the authors in \cite{chen2012new} improved the scheme in \cite{hohenberger2005securely}
in both efficiency and verifiability. Then, the authors in \cite{wang2014securely}
made further improvement by reducing the two servers model to one
single untrusted server model. In the following, we will compare our
ExpSOS with the three schemes in \cite{hohenberger2005securely,chen2012new,wang2014securely}.

In both \cite{hohenberger2005securely} and \cite{chen2012new}, the
authors consider outsourcing modular exponentiation to two untrusted
servers $S_{1}$ and $S_{2}$ and it is assumed that the two servers
do not collude which corresponds to our MM model. In both schemes,
a subroutine $\mathsf{Rand}$ is utilized to generate random modular
exponentiation pairs. Specifically, on input a base $g\in\mathbb{Z}_{p}^{*}$,
the subroutine $\mathsf{Rand}$ will generate random pairs in the
form of $(\theta,g^{\theta}\mod p)$, where $\theta$ is a random
number in $\mathbb{Z}_{p}^{*}$. Then the end-user can make queries
to $\mathsf{Rand}$ and each query will return a random pair to the
end-user. Typically, the subroutine $\mathsf{Rand}$ is implemented
via two different methods. One method is that a table of random pairs
is pre-computed from a trusted server and stored at the end-user.
Whenever the end-user needs to make a query to $\mathsf{Rand}$, it
just randomly draw a pair from the table. The critical problem of
this method is that it will take a lot of storage space from the end-user.
Specifically, a random pair will take $2l_{p}$ space, where $l_{p}$
is the bit length of $p$. In addition, to make the generation of
the pairs look random, the table size should be large. As a result,
the storage overhead becomes unacceptable for the resource-constrained
end-users. The other method is to utilize some pre-processing techniques
such as the $\mathsf{BPV}$ generator \cite{boyko1998speeding} and
the the $\mathsf{EBPV}$ generator \cite{nguyen2001distribution}.
To generate one random pair, the $\mathsf{EBPV}$ generator takes
$\mathcal{O}(\log^{2}l_{a})$ modular multiplications, where $l_{a}$
is the bit length of the exponent. 

The scheme proposed in \cite{hohenberger2005securely} can be briefly
summarized as follows. First, the end-user runs $\mathsf{Rand}$ $6$
times to obtain random pairs $(\alpha,g^{\alpha}),(\beta,g^{\beta}),(t_{1},g^{t_{1}}),$
$(t_{2},g^{t_{2}}),(r_{1},g^{r_{1}}),(r_{2},g^{r_{2}})$. Then $u^{\alpha}$
can be written as 
\[
u^{a}=v^{b}f^{a-b}\left(\frac{v}{f}\right){}^{a-b}\left(\frac{u}{v}\right){}^{d}\left(\frac{u}{v}\right){}^{a-d},
\]
where $v=g^{\alpha},b=\frac{\beta}{\alpha}$, $f$ and $d$ are random
integers. The end-user then makes queries in random order to the cloud
server $S_{1}$ $Q_{1}^{1}=\left(\frac{u}{v}\right){}^{d},Q_{1}^{2}=f{}^{a-b},Q_{1}^{3}=\left(g^{r_{1}}\right){}^{\frac{t_{1}}{r_{1}}},Q_{1}^{4}=\left(g^{r_{2}}\right){}^{\frac{t_{2}}{r_{2}}}$.
Similarly, the end-user makes queries to the second cloud server $S_{2}$
$Q_{2}^{1}=\left(\frac{u}{v}\right){}^{a-d},Q_{2}^{2}=\left(\frac{v}{f}\right){}^{a-b},Q_{2}^{3}=\left(g^{r_{1}}\right){}^{\frac{t_{1}}{r_{1}}},Q_{2}^{4}=\left(g^{r_{2}}\right){}^{\frac{t_{2}}{r_{2}}}$.
The result can be recovered as $u^{a}=g^{\beta}\cdot Q_{1}^{1}\cdot Q_{1}^{2}\cdot Q_{2}^{1}\cdot Q_{2}^{2}$.
The result verification is carried out by checking whether $Q_{1}^{3}=Q_{2}^{3}=g^{t_{1}}$
and $Q_{1}^{4}=Q_{2}^{4}=g^{t_{2}}$. We note that the end-user needs
to make queries to each server $S_{1}$ and $S_{2}$ for four times,
among which the first two are computation queries and the other two
are test queries. Since the test queries and the computation queries
are independent, the servers can potentially compute the test queries
honestly but cheat in the computation queries. The authors address
this problem by sending out the queries in random order. The verifiability
of this scheme is $\frac{1}{2}$. In the outsourcing process, $E$
has to run the subroutine $\mathsf{Rand}$ 6 times, make 9 modular
multiplications ($MMul$) and 5 modular inversions ($MInv$), where
$\mathsf{Rand}$ has a complexity of $\mathcal{O}(\log^{2}n)$ $MMul$
and $n$ is the bit length of the exponent.

Based on \cite{hohenberger2005securely}, the authors in \cite{chen2012new}
made some improvement by reducing the computational overhead to $5$
$\mathsf{Rand}$, $7$ $MMul$ and $3$$MInv$ and the queries to
the two servers are reduced to $6$ times in total. Moreover, the
verifiability is improved to $\frac{2}{3}$.

In comparison, our ExpSOS under MM model can be modified as in Algorithm
\ref{alg:CryptSOS-under-MM}. Since the cloud servers $S_{1}$ and
$S_{2}$ do not collude, the only way to make the equality condition
satisfied is that $S_{1}$ and $S_{2}$ both compute honestly. Thus
the verifiability is $1$. Moreover, in this process, we successfully
avoid inversion that is considered much more expensive than multiplication
in field operations. The total computational overhead is only 3 $MMul$. 

\begin{table*}
\caption{Numeric Results\label{tab:Numeric-Results}}

\centering{}%
\begin{tabular}{|c|c|c|c|c|c|c|c|c|c|}
\hline 
\multirow{3}{*}{$l_{N}$(bits)} & \multicolumn{9}{c|}{$l_{b}(bits)$}\tabularnewline
\cline{2-10} 
 & \multicolumn{3}{c|}{$4$} & \multicolumn{2}{c|}{$8$} & \multicolumn{2}{c|}{$12$} & \multicolumn{2}{c|}{$16$}\tabularnewline
\cline{2-10} 
 & $t_{0}$ ($ms$) & $t_{s}$ ($ms$) & $\tau$ & $t_{s}$ ($ms$) & $\tau$ & $t_{s}$ ($ms$) & $\tau$ & $t_{s}$ ($ms$) & $\tau$\tabularnewline
\hline 
$128$ & $1358$ & $87$ & $15.6$ & $216$ & $6.3$ & $321$ & $4.2$ & $397$ & $3.4$\tabularnewline
\hline 
$256$ & $2554$ & $89$ & $28.6$ & $244$ & $10.5$ & $346$ & $7.4$ & $459$ & $5.6$\tabularnewline
\hline 
$384$ & $4095$ & $127$ & $32.3$ & $249$ & $16.5$ & $358$ & $11.4$ & $463$ & $8.8$\tabularnewline
\hline 
$512$ & $7837$ & $134$ & $58.6$ & $281$ & $27.9$ & $399$ & $19.6$ & $496$ & $15.8$\tabularnewline
\hline 
$640$ & $10991$ & $146$ & $75.0$ & $288$ & $38.2$ & $423$ & $26.0$ & $627$ & $17.5$\tabularnewline
\hline 
$768$ & $11427$ & $148$ & $77.2$ & $295$ & $38.7$ & $433$ & $26.4$ & $642$ & $17.8$\tabularnewline
\hline 
$896$ & $17445$ & $158$ & $110.2$ & $317$ & $54.9$ & $451$ & $38.7$ & $680$ & $25.6$\tabularnewline
\hline 
$1024$ & $20235$ & $174$ & $116.2$ & $329$ & $61.5$ & $504$ & $40.1$ & $739$ & $27.4$\tabularnewline
\hline 
\end{tabular}
\end{table*}

\begin{algorithm}[tbh]
\caption{ExpSOS under MM Model\label{alg:CryptSOS-under-MM}}

\smallskip 
\textbf{Input:} $N,u,a\in\mathbb{R}_N$.\\
\textbf{Output:} $R_0=u^a\bmod N$, $\Lambda=\{\mathsf{True},\mathsf{False}\}.$

$\mathsf{Key~Generation}$:
\begin{algorithmic}[1]
\STATE $E$ generates a large prime number $p$ and calculate $L\leftarrow pN$. The public key is $K_{p}=\{L\}$ and the private key is $K_s=\{p,N\}$.
\STATE $E$ selects random integers $r,k\in\mathbb{Z}_N$ as the temporary key.
\end{algorithmic}

$\mathsf{Problem~Transformation}$
\begin{algorithmic}[1]
\STATE $E$ calculates $A\leftarrow a+k\phi(N)$ and $U\leftarrow (u+rN)\bmod L$. 
\STATE $E$ then outsources $\{U,A,L \}$ to both  cloud servers $S_1$ and $S_2$.
\end{algorithmic}

$\mathsf{Cloud Computation}$:
\begin{algorithmic}[1]
\STATE  $S_1$ computes $R_1\leftarrow U^A\bmod L$ and $S_2$ computes $R_2\leftarrow U^A\bmod L$. 
\STATE The results $R_1$ and $R_2$ are returned to $E$.
\end{algorithmic}

$\mathsf{Result~Verification}$
\begin{algorithmic}[1]
\STATE $E$ checks $R_1\bmod N=R_2\bmod N$. Set $\Lambda \leftarrow  \mathsf{True}$ if the equality holds; otherwise set $\Lambda \leftarrow  \mathsf{False}$.
\end{algorithmic}

$\mathsf{Result~Recovery}$:
\begin{algorithmic}[1]
\STATE $E$ recovers the result as $R \leftarrow  R_1\bmod N$.
\end{algorithmic}
\end{algorithm}

In \cite{wang2014securely}, the authors assume a Malicious Single
server (MS) model. Similarly, the scheme utilizes a subroutine $\mathsf{Rand}$
via some pre-processing techniques such as $\mathsf{BPV^{+}}$ that
is a modified version of $\mathsf{BPV}$. The scheme in \cite{wang2014securely}
can be summarized as follows. First, the end-user runs $\mathsf{Rand}$
$7$ times to obtain random pairs $(\alpha_{1},g^{\alpha_{1}})\ensuremath{,}(\alpha_{2},g^{\alpha_{2}})\ensuremath{,}(\alpha_{3},g^{\alpha_{3}})\ensuremath{,}(\alpha_{4},g^{\alpha_{4}})\ensuremath{,}(t_{1},g^{t_{1}})\ensuremath{,}(t_{2},g^{t_{2}}),$
$(t_{3},g^{t_{3}})$. Then it calculates $c=(a-b\chi)\bmod p\ensuremath{,}$
$\omega=u/\mu_{1}\ensuremath{,}h=u/\mu_{3},$ and $\theta=(\alpha_{1}b-\alpha_{2})\chi+(\alpha_{3}c-\alpha_{4})\bmod p$,
where $\chi,b$ are randomly selected and $\mu_{i}=g^{\alpha_{i}},$
for $i=1,2,3,4$. The end-user then queries to a single cloud server
$S$ $Q^{1}=\left(g^{t_{1}}\right){}^{\frac{\theta}{t_{1}}}\ensuremath{,}Q^{2}=\left(g^{t_{2}}\right){}^{\frac{t_{3}-\theta}{t_{2}}},Q^{3}=\omega^{b}\ensuremath{,}Q^{4}=h^{c}$.
The result is recovered as $u^{a}=(\mu_{2}\cdot Q^{3})^{\chi}\cdot Q^{1}\cdot\mu_{4}\cdot Q^{4}$.
The result verification is carried out by checking whether $Q^{1}\cdot Q^{2}=g^{t_{3}}$
is true. Similarly, the queries can be divided as test queries and
computation queries. As a consequence, the cloud can compute honestly
on the test queries and cheat on the computation queries. Thus, due
to the random order of the queries, the verifiability of this scheme
is $\frac{1}{2}$. We note that in the result recovery process, the
end-user has to compute an exponentiation $(\mu_{2}\cdot\omega^{b})^{\chi}$
which takes $\frac{3}{2}\log\chi$ multiplications. The whole scheme
will take $7$ $\mathsf{Rand}$, $12+\frac{3}{2}\log\chi$ $MMul$,
$4$ $MInv$ and make $4$ queries to the cloud server. In comparison,
ExpSOS can avoid inversion and only needs $(5+3\log b)$ $MMul$,
where $b$ is a small integer. 

In terms of security, we have shown that ExpSOS can successfully conceal
the base, exponent and the modulus of the modular exponentiation.
It is computationally infeasible for the cloud to derive any key information
from the disguised problem. In comparison, all the above three schemes
\cite{hohenberger2005securely,chen2012new,wang2014securely} can only
conceal the exponent and base while the modulus is exposed to the
cloud. Thus ExpSOS can provide much improved security. Moreover, the
three schemes in \cite{hohenberger2005securely},\cite{chen2012new}
and \cite{wang2014securely} achieve verifiability of $\frac{1}{2}$,
$\frac{2}{3}$ and $\frac{1}{2}$ respectively. In comparison, the
verifiability of ExpSOS is $1-\frac{1}{2b^{2}}$ that is close to
$1$. This means that the end-user is more confident about the results
returned by the cloud. Furthermore, the security of the schemes in
\cite{hohenberger2005securely} and \cite{chen2012new} relies on
the assumption that the two cloud servers will not collude. The scheme
\cite{wang2014securely} and our proposed ExpSOS are applicable to
one single untrusted server hence eliminating the non-collusion assumption.

The comparison of ExpSOS and the schemes in \cite{hohenberger2005securely,chen2012new,wang2014securely}
is summarized in Table \ref{tab:Performance-Comparison-1}. We can
see that our proposed ExpSOS outperforms other schemes in both computational
complexity and security. ExpSOS also makes the least queries to the
cloud that will introduce the least communication overhead. Moreover,
ExpSOS is cost-aware in computational overhead and security such that
the end-users can select the most suitable outsourcing scheme according
to their own constraints and demands. Also, ExpSOS can be modified
such that it is applicable to HCS, MM and MS model.

\subsection{Numeric Results}

In this section, we measure the performance of ExpSOS for modular
exponentiation through simulation in mobile phones. The computation
of both the end-user and the cloud server is simulated in the same
phone Samsung GT-I9100 with Android 4.1.2 operating system. The CPU
is Dual-core 1.2 GHz Cortex-A9 with $1$ GB RAM. In the outsourcing
process, we focus on the computational gain, denoted as $\tau$, from
the outsourcing. We measure the local processing time ($t_{0}$) to
compute the modular exponentiation $u^{a}\bmod N$ without outsourcing
and the local processing time ($t_{s}$) with outsourcing which includes
the problem transformation, result recovery and result verification.
To measure the performance of ExpSOS under different levels of complexity,
we let the size of the ring $l_{N}$ vary from $128$ bits to $1024$
bits. Also, to show the cost-awareness of ExpSOS, we let the size
of the security parameter $l_{b}$ vary from $4$ bits to $16$ bits.
The processing time is averaged over $1000$ independent rounds. The
numeric result is shown in Table \ref{tab:Numeric-Results} where
each number stands for the average processing time for $100$ rounds.
We can see that when the size of the ring $l_{N}$ increases, the
performance gain $\tau$ also increases for the same security parameter
$b$. This means that when the original problem is more complex, ExpSOS
would have a better performance. The reason is that the complexity
of modular exponentiation depends on the number of multiplications
that is positively correlated to the logarithm of the size of the
ring $l_{N}$. However, in ExpSOS the local processing takes almost
the same number of multiplications for a fixed security parameter
$b$. We can also see that there exists a trade-off between security
and computational overhead. When $b$ increases, the computational
overhead increases accordingly. Since the verifiability is $1-\frac{1}{2b^{2}}$,
a bigger $b$ means better security guarantees.

\section{Conclusion\label{sec:Conclusion}}

In this paper, we design a secure outsourcing scheme ExpSOS that can
be widely used to outsource general exponentiation operations for
cryptographic computations, including modular exponentiation and scalar
multiplication. The proposed ExpSOS enables end-users to outsource
the computation of exponentiation to a single untrusted server at
the cost of only a few multiplications. We also provide a verification
scheme such that the result is verifiable with probability $1-\frac{1}{2b^{2}}$.
With the security parameter $b$, ExpSOS is cost-aware in that it
can provide different security levels at the cost of different computational
overhead. The comprehensive evaluation demonstrates that our scheme
ExpSOS can significantly improves the existing schemes in efficiency,
security and result verifiability.

\end{document}